# A General Model of Interfacial Chemical Equilibrium in Phase-Field Method

Yanzhou Ji[1,2] and Long-Qing Chen[1]

[1]*Department of Materials Science and Engineering, The Pennsylvania State University, University Park, PA 16802, USA*

[2]*Department of Materials Science and Engineering, The Ohio State University, Columbus, OH 43210, USA*

## Abstract

We report a new approach to interfacial equilibria among multiple solution phases in the phase-field method. It employs auxiliary non-conserved variables to describe the composition differences among different phases, and their temporal evolution is driven by the differences among different chemical diffusional potentials. It is reduced to the Wheeler-Boettinger-McFadden (WBM) model of equal interfacial chemical compositions and to the Kim-Kim-Suzuki (KKS) model of equal interfacial chemical diffusion potentials as two limiting cases. It can directly incorporate thermodynamic databases without any further approximations and simplifications. It is generally applicable to a wide range of problems involving composition evolution and chemical equilibria, including processes such as interdiffusion between two ordered phases not sharing any common composition range, which would pose difficulty to the existing treatments using WBM and KKS models.

Emails: ji.730@osu.edu (Yanzhou Ji) and lqc3@psu.edu (Long-Qing Chen)

This manuscript is based on work completed in 2023. The work was previously submitted to a journal but was not published because the revision process was not completed. The present version has been updated and is released as an arXiv preprint.

## 1. Introduction

Many materials processes involve local composition changes among different phases in a microstructure. Notable examples include diffusional phase transformations such as atomic ordering and compositional phase separation, Ostwald ripening, solidification of alloys, corrosion/oxidation and other chemical reactions. Over the past few decades, the phase-field approach[1–8] has been extensively applied to predicting diffusional microstructure evolution. Based on the "diffuse-interface" description, the phase-field approach avoids the explicit tracking of interface positions by introducing gradient energy. Combining with thermodynamic and kinetic databases, it can be directly applied to realistic multicomponent multiphase materials of practical interests.

In the phase-field method of diffusional phase transformations, a phase interface, i.e., across which the phase order parameter smoothly varies, is often treated as a mixture of multiple phases. The interface compositions are treated as weighted sums of the compositions in different phases. However, this relation alone is not sufficient to determine the different phase compositions at interfaces, and additional constraints among the different phase compositions has to be imposed in phase-field models of diffusional phase transformations. For example, the earlier Wheeler-Boettinger-McFadden (WBM) model[9] assumes equal phase compositions at interfaces, whereas the widely adopted Kim-Kim-Suzuki (KKS) model[10] assumes equal diffusion chemical potential ($1^{st}$ derivative of free energy with respect to composition) at interfaces. In comparison to the WBM model, the KKS model removes the extra non-equilibrium chemical potential at interfaces, allowing for larger simulation size scales[10].

However, both existing models may pose difficulties in their practical implementations in phase-field simulations, and the assumptions in either of the models may deviate significantly from the realistic interfacial chemical states for specific processes. For example, the equal phase composition assumption in the WBM model works only if all the solution phases share some common composition ranges. A familiar example that the WBM model will not be applicable is the binary Al-Cu alloys[11] in which the non-stoichiometric θ phase is described by the sublattice model $Al_{2/3}(Al,Cu)_{1/3}$, indicating an allowable Cu composition range of [0,1/3], while the non-stoichiometric η phase is described by $(Al,Cu)_{1/2}Cu_{1/2}$, indicating a Cu composition range of [1/2,1]. As a result, there is no common Cu composition range between the two phases to satisfy the equal composition assumption at θ/η interfaces in the WBM model. On the other hand, solving the phase compositions from the equal diffusion potential condition in the KKS model requires the time-consuming solutions to nonlinear algebraic equations using iterative methods. Simplifying approximations of realistic free energy functions using parabolas[12,13] may cause other issues such as unphysical compositions (higher than 1 or lower than 0) and thus inaccurate growth kinetics. Furthermore, the interfacial composition profiles among different solution phases may deviate significantly from either the equal composition conditions or the equal diffusion potential conditions in practical diffusional processes.

In this work, we propose a general model of interfacial equilibria that can be easily implemented in existing phase-field models. It is reduced to the WBM and KKS models as two limiting cases. In particular, we introduce new auxiliary non-conserved variables to describe composition differences among different solution phases. These auxiliary variables evolve according to the Allen-Cahn equations[14] driven by the differences in the diffusion potentials for each component. Therefore, the evolution of interfacial compositions is described by a combination of the conserved composition fields and auxiliary variables. We demonstrate that the proposed model is general and flexible, including those under equal composition and equal diffusion potential condition as limiting cases. It is even applicable to systems involving chemical equilibria between two non-stoichiometric ordered phases without common composition ranges of existence. Furthermore, it can directly incorporate existing thermodynamic databases without any furthermore simplifications.

## 2. Model Description

For simplicity, as an example of illustration, we consider an A-B binary thermodynamic system consisting of two solution phases α and β. The two phases are distinguished by a continuously varying phase order parameter $\xi$ so that $\xi=0$ in the equilibrium α phase, $\xi=1$ in the equilibrium β phase, and $0<\xi<1$ in the α/β interfacial region.

The composition of each phase can be described by its respective atomic fraction of the diffusing species B, $X_B^{\varphi}$ $(\varphi = \alpha, \beta)$, with $X_A^{\varphi} = 1 - X_B^{\varphi}$. At the α/β interface, we consider the local composition field $X_B$ as a mixture of the composition of the two phases:

$$X_B = \left(1 - h(\xi)\right) X_B^{\alpha} + h(\xi) X_B^{\beta} \tag{1}$$

where $h(\xi) = 6\xi^5 - 15\xi^4 + 10\xi^3$ is the commonly used interpolation function in phase-field method. To uniquely determine $X_B^{\alpha}$ and $X_B^{\beta}$ for each given combination of $\xi$ and $X_B$, here we define an auxiliary parameter $\phi$ to describe the composition difference between the two phases:

$$\phi = X_B^\alpha - X_B^\beta \tag{2}$$

Therefore, the phase compositions can be expressed as a function of the phase order parameter $\xi$, the local composition field $X_B$, and the auxiliary variable $\phi$:

$$X_B^\alpha = X_B + h(\xi)\phi \tag{3a}$$

$$X_B^\beta = X_B - \big(1 - h(\xi)\big)\phi \tag{3b}$$

The total free energy $F$ of the system can be expressed as a functional of these phase-field variables. For simplicity, we only consider the local bulk free energy density $f_{bulk}$ and the interfacial energy density $f_{int}$, without considering the long-range elastic and electrostatic interactions:

$$F = \int [f_{bulk}(\xi, X_B, \phi, T) + f_{int}(\xi, \nabla\xi)]dV \tag{4}$$

where $V$ is volume, and $T$ is temperature. The local bulk free energy density $f_{bulk}$ includes the contribution from the composition-dependent phase chemical potentials $\mu^\alpha$ and $\mu^\beta$:

$$\begin{aligned} f_{bulk} &= \big(1 - h(\xi)\big)\frac{\mu^\alpha(X_B^\alpha, T)}{v_m} + h(\xi)\frac{\mu^\beta\left(X_B^\beta, T\right)}{v_m} \\ &= \big(1 - h(\xi)\big)\frac{\mu^\alpha(\xi, X_B, \phi, T)}{v_m} + h(\xi)\frac{\mu^\beta(\xi, X_B, \phi, T)}{v_m} \end{aligned} \tag{5}$$

where $\mu^\alpha$ and $\mu^\beta$ are chemical potentials of $\alpha$ and $\beta$ phases, and $v_m$ is the molar volume. $f_{int}$ is typically formulated using a double-well-type function $g(\xi) = \xi^2(1-\xi)^2$ and gradient energy terms:

$$f_{int} = wg(\xi) + \frac{1}{2}\kappa_\xi(\nabla\xi)^2 + \frac{1}{2}\kappa_X(\nabla X_B)^2 \quad (6)$$

where $\kappa_\xi$ and $\kappa_X$ are the gradient energy coefficients, which, together with the double-well height $w$, describes the interfacial energy and interface thickness. Note that here we keep the compositional gradient term $\frac{1}{2}\kappa_X(\nabla X_B)^2$ to account for possible phase separations described by the Cahn-Hilliard equation.

The evolution of the non-conserved phase order parameters $\xi$ is governed by the Allen-Cahn equation[14], whereas the evolution of the conserved composition field $X_B$ generally follows the Cahn-Hilliard equation[15]:

$$\frac{\partial\xi}{\partial t} = -L_\xi\frac{\delta F}{\delta\xi} = -L_\xi\left(wg'(\xi) - \kappa_\xi\nabla^2\xi + \frac{\partial h}{\partial\xi}\frac{\mu^\beta - \mu^\alpha}{v_m} + \frac{1-h(\xi)}{v_m}\frac{\partial\mu^\alpha}{\partial\xi} + \frac{h(\xi)}{v_m}\frac{\partial\mu^\beta}{\partial\xi}\right) \quad (7a)$$

$$\frac{\partial X_B}{\partial t} = \nabla\cdot M_X\nabla\frac{\delta F}{\delta X_B} = \nabla\cdot M_X(\xi)\nabla\left(\frac{1-h(\xi)}{v_m}\frac{\partial\mu^\alpha}{\partial X_B} + \frac{h(\xi)}{v_m}\frac{\partial\mu^\beta}{\partial X_B} - \kappa_X\nabla^2 X_B\right) \quad (7b)$$

where $L_\xi$ is the kinetic coefficient related to interface mobility, $M_X(\xi)$ is the phase-dependent interdiffusion mobility. Similarly, we can write down the evolution equation of the auxiliary variable $\phi$ following the Allen-Cahn form:

$$\frac{\partial\phi}{\partial t} = -L_\phi\frac{\delta F}{\delta\phi} = -L_\phi\left(\frac{1-h(\xi)}{v_m}\frac{\partial\mu^\alpha}{\partial\phi} + \frac{h(\xi)}{v_m}\frac{\partial\mu^\beta}{\partial\phi}\right) = -\frac{L_\phi}{v_m}h(\xi)\big(1-h(\xi)\big)\left(\frac{\partial\mu^\alpha}{\partial X_B^\alpha} - \frac{\partial\mu^\beta}{\partial X_B^\beta}\right) \quad (8)$$

where $L_\phi$ is the relaxation parameter for the evolution of $\phi$. Please refer to Supplementary Materials S1 for the detailed derivations. This equation can guarantee that the equal diffusion potential condition $\tilde{\mu}_B = \tilde{\mu}_B^{\alpha} = \tilde{\mu}_B^{\beta} = \frac{\partial \mu^{\alpha}}{\partial X_B^{\alpha}} = \frac{\partial \mu^{\beta}}{\partial X_B^{\beta}}$ is automatically achieved for $\phi$ at equilibrium ($\frac{\partial \phi}{\partial t} = 0$) at the interfaces. However, this equation could not guarantee that the equal diffusion potential condition is achieved in the bulk of α and β away from the interface, because the term $h(\xi)\big(1 - h(\xi)\big)$ is always zero in the bulk. Therefore, we propose to further improve the governing equation of $\phi$ by removing the $h(\xi)\big(1 - h(\xi)\big)$ term:

$$\frac{\partial \phi}{\partial t} = -\frac{L_{\phi}}{v_m}\left(\frac{\partial \mu^{\alpha}}{\partial X_B^{\alpha}} - \frac{\partial \mu^{\beta}}{\partial X_B^{\beta}}\right) \tag{9}$$

This equation is more straight-forward in the sense that it becomes a relaxational form of the KKS model: if Eq. (9) is allowed to evolve much faster than $\xi$ and $X_B$, e.g., using a large $L_{\phi}$ value or iterating the equation multiple times within every evolution step of $\xi$ and $X_B$, the equal diffusion potential condition of $\frac{\partial \mu^{\alpha}}{\partial X_B^{\alpha}} = \frac{\partial \mu^{\beta}}{\partial X_B^{\beta}}$ can be achieved in the whole system at each evolution step of $\xi$ and $X_B$. Moreover, it would not change the condition of equal diffusion potential at interfaces at equilibrium. Numerical solutions to these governing equations yield the temporal evolution of the composition profiles and thus the microstructure evolution.

For comparison, we summarize the formulations of the WBM model, the KKS model and the current model for this binary two-phase system in Table 1. In contrast to the WBM and KKS models, the current model allows one to model the full range of interfacial conditions from the equal composition condition in the WBM model to the equal diffusion potential condition in the KKS model. For example, the limiting case of equal interfacial composition assumption in the WBM model corresponds to the special case with $\phi = 0$ and $L_\phi = 0$ in the present model. On the other hand, the limiting case of equal diffusion potential assumption in the KKS model corresponds to the special case where $\phi$ evolves to equilibrium to ensure $\tilde{\mu}_B = \tilde{\mu}_B^\alpha = \tilde{\mu}_B^\beta = \frac{\partial \mu^\alpha}{\partial X_B^\alpha} = \frac{\partial \mu^\beta}{\partial X_B^\beta}$ at each evolution step of $\xi$ and $X_B$, and that the extra potential $\mu_{ex} = (1 - h(\xi))\mu^\alpha(X_B^\alpha) + h(\xi)\mu^\beta(X_B^\beta) - (1 - h(\xi^{eq}))\mu^\alpha(X_B^{\alpha,eq}) - h(\xi^{eq})\mu^\beta(X_B^{\beta,eq})$ is removed at interfaces. However, the more common cases in practice are likely between the two extreme cases of equal chemical composition interfacial condition described by the WBM model and equal diffusion potential interfacial condition described by the KKS model. Such cases can be rather naturally modeled by the proposed model as illustrated in Figure 1.

Table 1: Formulations of the WBM model, the KKS model and the current model.

| | WBM model | KKS model | Current model |
|---|---|---|---|
| Interfacial composition | $X_B = X_B^\alpha = X_B^\beta = (1 - h(\xi))X_B^\alpha + h(\xi)X_B^\beta$ | $X_B = (1 - h(\xi))X_B^\alpha + h(\xi)X_B^\beta$, $X_B^\alpha \neq X_B^\beta$ | $X_B = (1 - h(\xi))X_B^\alpha + h(\xi)X_B^\beta$ |
| Diffusion potential | $\tilde{\mu}_B = (1 - h(\xi))\tilde{\mu}_B^\alpha + h(\xi)\tilde{\mu}_B^\beta$, $\tilde{\mu}_B^\alpha \neq \tilde{\mu}_B^\beta$ | $\tilde{\mu}_B = \tilde{\mu}_B^\alpha = \tilde{\mu}_B^\beta = (1 - h(\xi))\tilde{\mu}_B^\alpha + h(\xi)\tilde{\mu}_B^\beta$ | $\tilde{\mu}_B = (1 - h(\xi))\tilde{\mu}_B^\alpha + h(\xi)\tilde{\mu}_B^\beta$ |
| Free energy expression | $F = \int_V \left( (1 - h(\xi)) \frac{\mu^\alpha(X_B)}{v_m} + h(\xi) \frac{\mu^\beta(X_B)}{v_m} + wg(\xi) + \frac{1}{2}\kappa_\xi (\nabla\xi)^2 + \frac{1}{2}\kappa_X (\nabla X_B)^2 \right) dV$ | $F = \int_V \left( (1 - h(\xi)) \frac{\mu^\alpha(X_B^\alpha)}{v_m} + h(\xi) \frac{\mu^\beta(X_B^\beta)}{v_m} + wg(\xi) + \frac{1}{2}\kappa_\xi (\nabla\xi)^2 \right) dV$ | $F = \int_V \left( (1 - h(\xi)) \frac{\mu^\alpha(X_B, \phi, \xi)}{v_m} + h(\xi) \frac{\mu^\beta(X_B, \phi, \xi)}{v_m} + wg(\xi) + \frac{1}{2}\kappa_\xi (\nabla\xi)^2 \right) dV$ |
| Evolution equation for $\xi$ | $\frac{\partial \xi}{\partial t} = -L_\xi \left( wg'(\xi) - \kappa_\xi \nabla^2 \xi + \frac{\partial h}{\partial \xi} \frac{\mu^\beta - \mu^\alpha}{v_m} \right)$ | $\frac{\partial \xi}{\partial t} = -L_\xi \left( wg'(\xi) - \kappa_\xi \nabla^2 \xi + \frac{\partial h}{\partial \xi} \frac{\mu^\beta - \mu^\alpha - \tilde{\mu}_B (X_B^\beta - X_B^\alpha)}{v_m} \right)$ | Eq. (7a) |
| Evolution equation for $X_B$ | $\frac{\partial X_B}{\partial t} = \nabla \cdot M_X(\xi) \nabla \left( \frac{1-h(\xi)}{v_m} \frac{\partial \mu^\alpha}{\partial X_B} + \frac{h(\xi)}{v_m} \frac{\partial \mu^\beta}{\partial X_B} - \kappa_X \nabla^2 X_B \right)$ | $\frac{\partial X_B}{\partial t} = \nabla \cdot M_X(\xi) \nabla \tilde{\mu}_B$ | Eq. (7b) |
| Additional treatment | Need to evaluate $\kappa_X$ according to the extra potential $\mu_{ex}$ | Need to solve $X_B^\alpha$ and $X_B^\beta$ from $\begin{cases} X_B = (1-h(\xi))X_B^\alpha + h(\xi) X_B^\beta \\ \frac{\partial \mu^\alpha}{\partial X_B^\alpha} = \frac{\partial \mu^\beta}{\partial X_B^\beta} \end{cases}$ | Need to evolve $\phi = X_B^\alpha - X_B^\beta$ according to Eq. (8) or Eq. (9) |

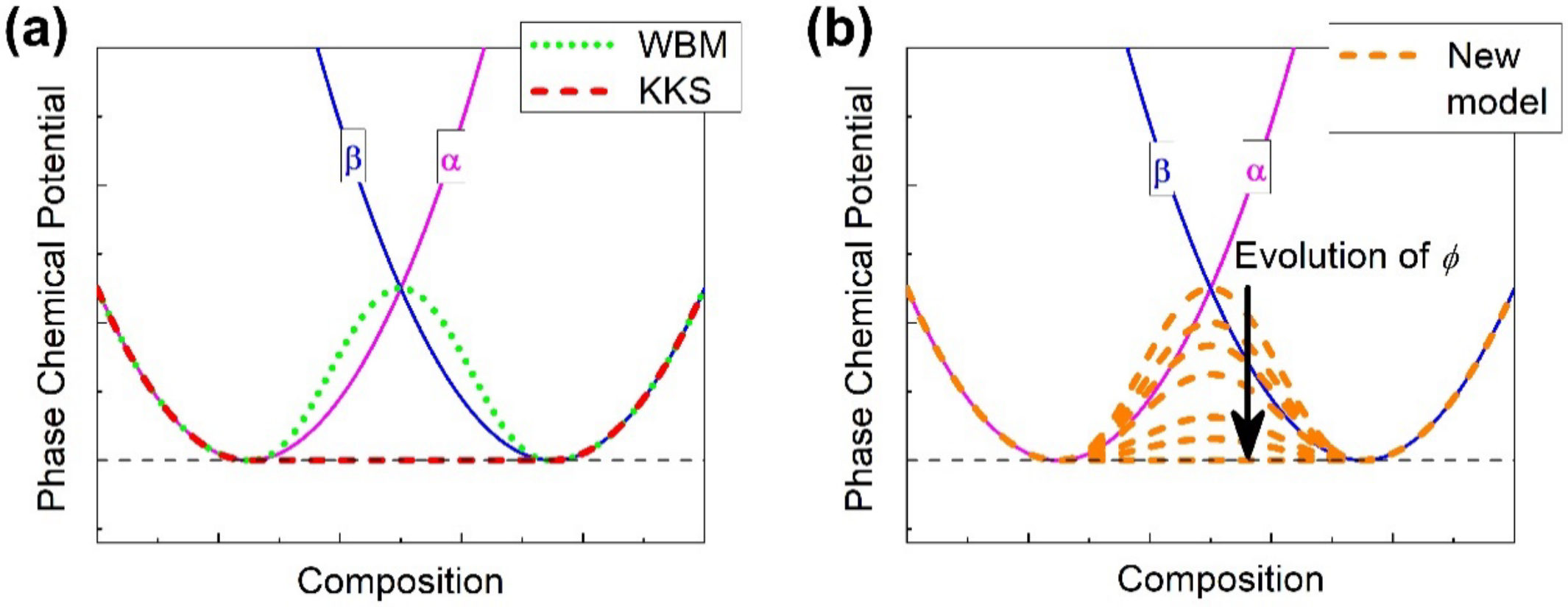

Figure 1: Illustration of (a) the WBM model, the KKS model and (b) the current phase-field model in terms of extra potential $\mu_{ex}$ at the interface. The magenta and blue lines represent the phase chemical potentials $\mu^\alpha$ and $\mu^\beta$ as a function of composition, respectively; the black dashed lines represent the common tangent line.

## 3. Results and Discussions

In this section, we present two application examples of our proposed model: (1) diffusional growth of α phase in β matrix of Ti-V alloys, where both α and β phases are fully disordered solid solutions; (2) interdiffusion between two nonstoichiometric phases, θ and η1, in Al-Cu alloys, without common composition range. We will also discuss the further extension of our current model.

### *3.1 Diffusional growth of α phase in Ti-V alloys*

To illustrate the application of the present model, we first consider the case of diffusional growth of a fully disordered phase from another supersaturated disordered phase where both the WBM and the KKS models are applicable. We choose the binary Ti-V alloy as a model system, in which both α-Ti ($\xi$=1) and β-Ti ($\xi$=0) are fully disordered solid solutions with complete composition range of [0, 1], as shown by the phase chemical potentials in Figure 2(a) from database[16].

For bench-marking purpose, we consider the diffusional growth of α-Ti from supersaturated β-Ti at $T$=873 K in a one-dimensional (1-D) system discretized using 1024 grid points and a relatively small grid spacing of $\Delta x$=1 nm. At this temperature, neither phases will phase separate, so the compositional gradient term $\frac{1}{2}\kappa_V(\nabla X_V)^2$ is neglected for the KKS model and our current model. An initial α-Ti nucleus with a length of 12 nm and composition of $X_V^{\alpha,0} = 0.0323$ was manually introduced at the left-hand-side of the 1-D system. The initial composition of the β matrix is $X_V^{\beta,0} = 0.1$, and the initial value of the compositional order parameter is set as $\phi = 0.0676$. The expressions for the phase chemical potentials (or the molar Gibbs free energies, or simply Gibbs energy, in most of the existing literature) $\mu^{\alpha}$ and $\mu^{\beta}$ are obtained from thermodynamic database[16]. We used a molar volume $v_m$ of $1.04 \times 10^{-5}$ m$^3$/mol. The gradient energy coefficient $\kappa_\xi$ is $4.09 \times 10^{-9}$ J/m, and the barrier height $w$ of the double-well potential is $1.1 \times 10^{9}$ J/m$^3$, which yields an interfacial energy of 0.5 J/m$^2$ and an interface thickness of 6 nm. The chemical mobility $M_X(\xi)$ is assumed to be a constant $2.61 \times 10^{-26}$ m$^5$/J/s for simplicity. The kinetic coefficient $L_\xi$ is $4.95 \times 10^{-8}$ m$^3$/J/s, and the kinetic coefficient $L_\phi$ ranges from 0 to $100L_\xi$. The same set of simulations were also performed using the WBM model with $\kappa_X = 1.84 \times 10^{-8}$ J/m and the KKS model for comparison. A modified Newton-Raphson method (see Supplementary Materials S2) was used to obtain the phase compositions $X_V^\alpha$ and $X_V^\beta$ in the KKS model. The test simulation results are summarized in Figure 2.

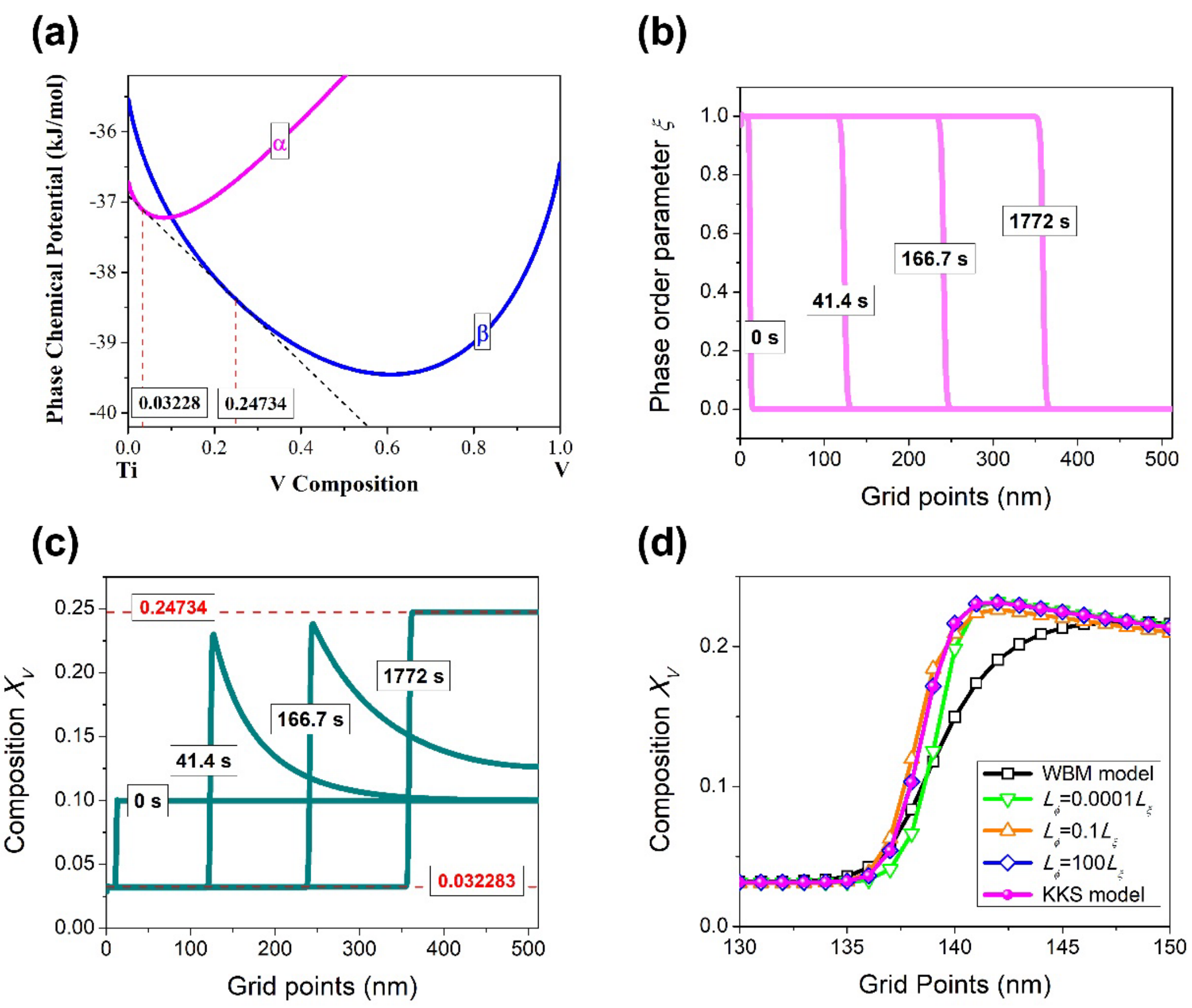

Figure 2: Phase-field simulation of diffusional growth of α-Ti from a supersaturated β-Ti matrix in a binary Ti-V alloy at 873 K, using the WBM model, the KKS model and the current phase-field models. (a) The phase chemical potential $\mu^\alpha$ and $\mu^\beta$ as a function of V composition. The black dashed line is the common tangent line. The equilibrium compositions are $X_V^{\alpha,eq} = 0.03228$ and $X_V^{\beta,eq} = 0.24734$. (b) Simulated $\xi$ profiles at four different growth times: $t$=0 s, 41.4 s, 166.7 s and 1772 s, using the current phase-field model with $L_\phi = 100L_\xi$. (c) Simulated

profiles of $X_V$ at the four different growth times using the current phase-field model with $L_\phi = 100L_\xi$. (d) Comparison of $X_V$ profiles at $t$=41.4 s using the WBM model, the KKS model and the current model with different $L_\phi$ values, $L_\phi = 0.0001L_\xi, L_\phi = 0.1L_\xi$ and $L_\phi = 100L_\xi$.

As shown in Figure 2, simulations using the current model converge to the correct equilibrium compositions in both α and β phases (Figure 2(c)). Note that the composition and order parameters profiles obtained from the WBM model, the KKS model, and the current models are very close to each other, so only the profiles from the current model with $L_\phi = 100L_\xi$ are shown in Figure 2(b-c). This indicates that the current phase-field model can well reproduce the predictions from both the KKS and the WBM models, and that the specific choice of the phase-field model (WBM, KKS and current model) would not significantly affect the α-Ti growth kinetics predictions in the current example with thin interfaces. However, as shown in Figure 2(d), during growth, the composition profiles at the two-phase interface are different for different models. In the current model, while $L_\phi = 0$ produces the same results as the WBM model, the composition profile approaches to that of the KKS model as $L_\phi$ increases.

To examine the differences among these models, we extract the profiles of $\phi$, diffusion potentials and extra potentials from different phase-field models. We also evolve $\phi$ using both Eq. (8) and Eq. (9) to show their differences. As shown in Figure 3(a), if $\phi$ is evolved using Eq. (8), the profile of $\phi$ would lie between the WBM (black line, i.e., $\phi=0$) and the KKS (magenta line) predictions. With the increase in $L_\phi$, $\phi$ would gradually approach to the KKS prediction near the α/β interface (at around 138 nm), while still significantly deviates from both predictions in the bulk, especially inside β. The profiles of the absolute diffusion potential difference in the two phases ($\left|\tilde{\mu}_V^\beta - \tilde{\mu}_V^\alpha\right|$) in Figure 3(b) also show the similar trend. As a result, the extra potential $\mu_{ex}$ can be largely reduced compared with the WBM model but may not be fully eliminated at the interface, as shown in Figure 3(c).

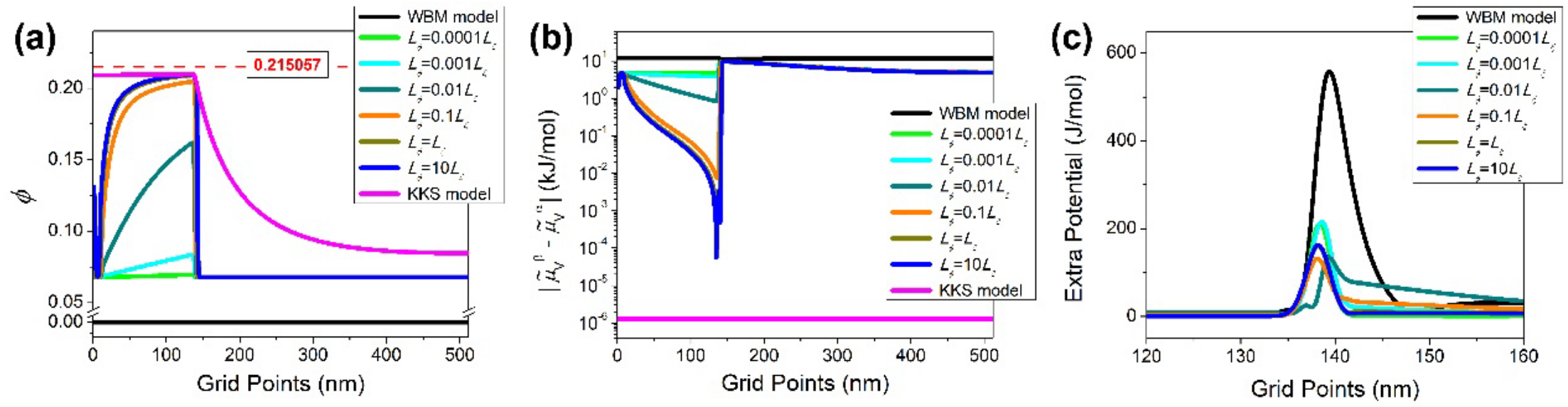

Figure 3: Comparison between the WBM model, the KKS model and the current phase-field models with Eq. (8) at $t$=41.4 s. (a), (b) and (c) show the $\phi$ profiles, the absolute difference of the diffusion potentials, and the extra potential from the current phase-field model with varying $L_\phi$ values, respectively. The red dashed lines in (a) indicate the equilibrium $\phi$ value of 0.215057.

In comparison, if $\phi$ is evolved using Eq. (9), with the increase in $L_\phi$, both $\phi$ and $\left|\tilde{\mu}_V^\beta - \tilde{\mu}_V^\alpha\right|$ would approach the KKS predictions throughout the whole system, as shown in Figures 4(a) and 4(b). The extra potential $\mu_{ex}$ is also more significantly reduced in Figure 4(c) than in Figure 3(c) under the same $L_\phi$ value. This difference lies in the removal of the $h(\xi)\left(1 - h(\xi)\right)$ term in Eq. (9), which can guarantee the evolution towards $\frac{\partial \mu^\alpha}{\partial X_B^\alpha} = \frac{\partial \mu^\beta}{\partial X_B^\beta}$ throughout the system, rather than only at interfaces. Therefore, although Eq. (8) is thermodynamically more consistent, Eq. (9) is more effective in reducing the extra potentials at interfaces.

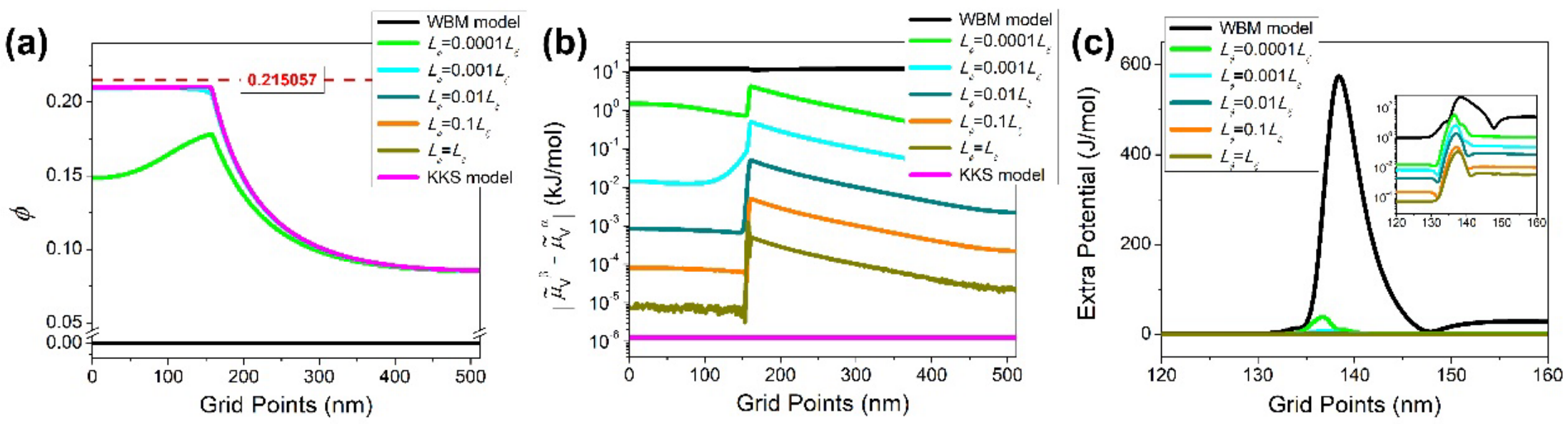

Figure 4: Comparison between the WBM model, the KKS model and the current phase-field models with Eq. (9) at $t$=41.4 s. (a), (b) and (c) show the $\phi$ profiles, the absolute difference of the diffusion potentials, and the extra potential from the current phase-field model with varying $L_\phi$ values, respectively. The red dashed lines in (a) indicate the equilibrium $\phi$ value of 0.215057. The inset in (c) shows the extra potential profiles in logarithm scale.

The simulation results in Figures 2-4 confirm the accuracy and effectiveness of the current phase-field model in predicting diffusional growth kinetics of precipitates. To further demonstrate the efficiency of the current phase-field model, we have performed 3-D simulations with larger grid spacing (10 nm) and interface thickness (60 nm), as shown in Supplementary Materials S3. The simulation results show that the current phase-field model can guarantee that more than 99.99% of the extra potential is removed at the interface during the evolution, while the simulation is ~3.3 times faster than the KKS model. This efficiency upgrade is more significant for 3-D systems than 1-D or 2-D systems, since there are more interface points in 3-D systems where more Newton iteration steps are needed for the KKS model.

### *3.2 Interdiffusion between nonstoichiometric phases in Al-Cu alloys*

It should be emphasized that the current model can be employed to model systems for which the application of existing WBM and/or KKS models may be challenging. As an example, we apply the current model to the interdiffusion between two nonstoichiometric phases without common compositional ranges. For simplicity, we consider the 1-D interdiffusion between the ordered θ and η1 phases in binary Al-Cu alloys at $T$=853 K. As shown in Table 2 and Figure 5(a), these two ordered phases do not share any common Cu composition range, yet the common tangent line can still be constructed between them. Therefore, if the two ordered phases get in contact while they both deviate from their equilibrium composition predicted by the common tangent line, interdiffusion and interface movement are expected to take place until the system reaches equilibrium. It should be emphasized again that the WBM model could not be applied here since the two phases have no common composition ranges. The application of KKS model to this example is possible but challenging, since the equal diffusion potential condition is difficult to be strictly achieved due to the extremely narrow composition ranges of θ with positive diffusion potentials and η1 with negative diffusion potentials. However, this challenging problem presents no difficulty to the current model. The 1-D system is discretized into 1024 grids with a grid spacing of $\Delta x$=1 nm. Initially, the left half of the system is the η1 phase ($\xi$=1) with composition $X_{Cu}^{\eta 1,0} = 0.55$ while the right half of the system is the θ phase ($\xi$=0) with composition $X_{Cu}^{\theta,0} = 0.3$. The simulation parameters are $v_m = 1.06 \times 10^{-5}$ $m^3$/mol, $\kappa_\xi = 4.09 \times 10^{-9}$ J/m, $w = 1.1 \times 10^9$ J/$m^3$ (corresponding to an interfacial energy of 0.5 J/$m^2$ and an interface thickness of 6 nm), $M_X(\xi) = 1.77 \times 10^{-24}$ $m^5$/J/s, $L_\xi = 2.81 \times 10^{-4}$ $m^3$/J/s, and $L_\phi = 2.81 \times 10^{-5}$ $m^3$/J/s. Here we emphasize that the chemical potentials of θ and η1 are directly taken from thermodynamic database[11] without doing any approximations, extrapolations and fitting. To avoid unphysical composition values out of the composition ranges during simulation, rather than directly evolve $X_{Cu}$ and $\phi$, we evolve two intermediate variables:

$$Y_1 = \ln \frac{3X_{Cu}^{\theta}}{1 - 3X_{Cu}^{\theta}} = \ln \frac{3[X_{Cu} + h(\xi)\phi]}{1 - 3[X_{Cu} + h(\xi)\phi]} \tag{10a}$$

$$Y_2 = \ln \frac{2X_{Cu}^{\eta 1} - 1}{2 - 2X_{Cu}^{\eta 1}} = \ln \frac{2\big[X_{Cu} - \big(1 - h(\xi)\big)\phi\big] - 1}{2 - 2\big[X_{Cu} - \big(1 - h(\xi)\big)\phi\big]} \tag{10b}$$

Therefore, we can obtain the expressions of $X_{Cu}$ and $\phi$ as a function of $\xi$, $Y_1$ and $Y_2$:

$$X_{Cu} = \left(1 - h(\xi)\right)\frac{e^{Y_1}}{3 + 3e^{Y_1}} + h(\xi)\frac{2e^{Y_2} + 1}{2 + 2e^{Y_2}} \tag{11a}$$

$$\phi = \frac{e^{Y_1}}{3 + 3e^{Y_1}} - \frac{2e^{Y_2} + 1}{2 + 2e^{Y_2}} \tag{11b}$$

Taking time derivatives on both sides of Eq. (11), we can obtain the governing equations for $Y_1$ and $Y_2$ (see Supplementary Materials S4 for the detailed derivations):

$$\frac{3e^{Y_1}}{(3 + 3e^{Y_1})^2}\frac{\partial Y_1}{\partial t} = \nabla \cdot M_{Cu}\nabla\frac{\delta F}{\delta X_{Cu}} - h(\xi)\frac{L_\phi}{v_m}\left(\frac{\partial \mu^\theta}{\partial X_{Cu}^\theta} - \frac{\partial \mu^{\eta 1}}{\partial X_{Cu}^{\eta 1}}\right) + \phi\frac{\partial h}{\partial \xi}L_\xi\frac{\delta F}{\delta \xi} \tag{12a}$$

$$\frac{2e^{Y_2}}{(2 + 2e^{Y_2})^2}\frac{\partial Y_2}{\partial t} = \nabla \cdot M_{Cu}\nabla\frac{\delta F}{\delta X_{Cu}} + \left(1 - h(\xi)\right)\frac{L_\phi}{v_m}\left(\frac{\partial \mu^\theta}{\partial X_{Cu}^\theta} - \frac{\partial \mu^{\eta 1}}{\partial X_{Cu}^{\eta 1}}\right) + \phi\frac{\partial h}{\partial \xi}L_\xi\frac{\delta F}{\delta \xi} \tag{12b}$$

The evolved variables $Y_1$ and $Y_2$ are then converted to $X_{Cu}$ and $\phi$ via Eq. (11).

Table 2: Information of θ and η1 phases in Al-Cu[11]

| Phases | Space group and prototype | Thermodynamic models | Cu composition range |
|---|---|---|---|
| θ | I4/mcm, $Al_2Cu$ | $(Al)_2(Al, Cu)_1$ | [0, 1/3] |
| η1 | C2/m, AlCu | $(Al, Cu)_1(Cu)_1$ | [1/2, 1] |

As shown in Figure 5, the initially non-equilibrium θ and η1 phases get in contact and reaches equilibrium via interdiffusion. The 1-D simulation using the new model correctly predicts the equilibrium compositions and compositional order parameter, as seen by the overlapping of the simulated composition and order parameter profiles (in solid lines) with the corresponding equilibrium values from thermodynamic calculations (in dashed lines). The θ/η1 interface moves towards the θ side due to the initially larger composition deviation from the equilibrium value on the η1 side. This example demonstrates the capability of our current phase-field model to deal with the interdiffusion between two nonstoichiometric phases that do not necessarily share any common composition range. In addition to the interdiffusion between two intermetallic phases, the present model is applicable to other practically important challenging examples include the interdiffusion and reaction between two nonstoichiometric oxides such as cathodes and electrolytes of solid oxide fuel cells (SOFCs).

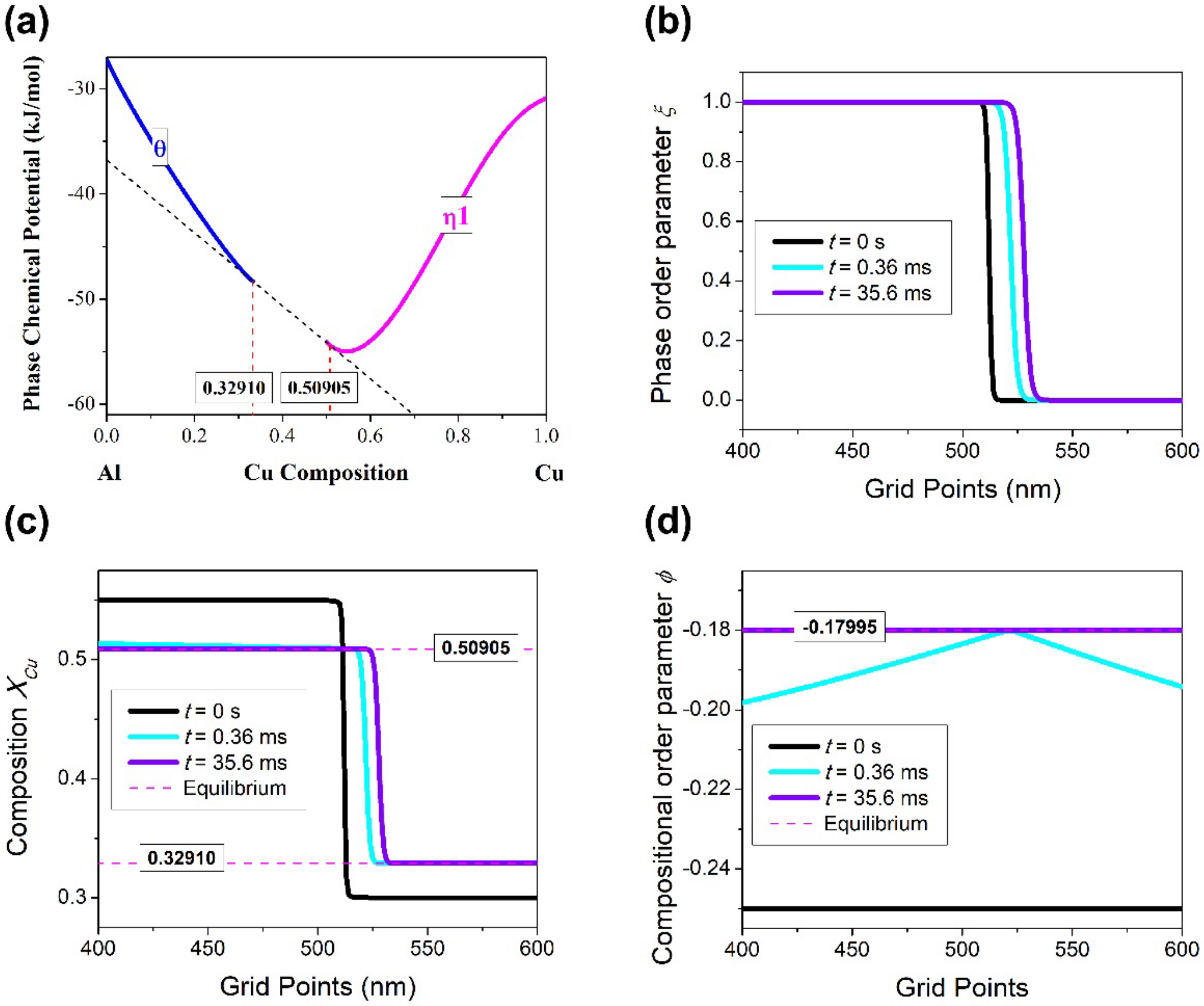

Figure 5: Simulation of interdiffusion between two nonstoichiometric phases θ and η1 in binary Al-Cu alloys using the current phase-field model. (a) Landscapes of $\mu^{\theta}$ and $\mu^{\eta 1}$ as a function of Cu composition from thermodynamic database. The black dashed line is the common tangent line. The equilibrium compositions are $X_{Cu}^{\theta,eq} = 0.32910$ and $X_{Cu}^{\eta 1,eq} = 0.50905$. (b) Simulated $\xi$ profiles at three different time steps: $t$=0 s, 0.36 ms, and 35.6 ms. (c) Simulated profiles of $X_{Cu}$ at the three different time steps. The equilibrium compositions are also shown by the red dashed lines. (d) Simulated $\phi$ profiles using Eq. (7b) at the three different time steps. The red dashed line indicates the equilibrium $\phi$ value of -0.17995.

### *3.3 Further extensions of the current model*

It should be emphasized that the current model can be easily extended to general multicomponent multiphase systems by defining the auxiliary variable for each diffusing species and between different solution phases. Consider a system with *M* solution phases and *N* independent compositions (the total number of compositions is *N*+1). For each of the *N* composition, we can define *M*-1 auxiliary variables characterizing the differences in phase compositions. Then there will be $N \times (M-1)$ auxiliary variables in total. For example, in the ternary Ni-Fe-Cr system[17], there are three solution phases at *T*=1673 K with complete composition range of [0, 1]: the liquid phase (liq), the α phase with the BCC structure and the γ phase with the FCC structure. In this system, there are *N*=2 independent compositions (e.g., Fe and Cr compositions) and *M*=3 phases. We can therefore define the following $N \times (M-1) = 4$ auxiliary variables:

$$\phi_1 = X_{Fe}^{\alpha} - X_{Fe}^{\gamma} \tag{13a}$$

$$\phi_2 = X_{Cr}^{\alpha} - X_{Cr}^{\gamma} \tag{13b}$$

$$\phi_3 = X_{Fe}^{liq} - X_{Fe}^{\gamma} \tag{13c}$$

$$\phi_4 = X_{Cr}^{liq} - X_{Cr}^{\gamma} \tag{13d}$$

The current model is also applicable to nonstoichiometric ordered phases described by sublattice models, as long as they share common compositions. For example, the lanthanum strontium manganite (LSM) perovskite is a nonstoichiometric phase usually used as SOFC cathodes, whose sublattice model is $(La^{3+}, Sr^{2+}, Mn^{3+}, Va)_1(Mn^{2+}, Mn^{3+}, Mn^{4+}, Va)_1(O^{2-}, Va)_3$[18]. The LSM phase has three independent compositions[19], e.g., La, Sr and Mn compositions. During operation, cation interdiffusion[20] can take place between LSM and SOFC electrolytes such as the yttrim-stabilized zirconia (YSZ), which is also a nonstoichiometric phase described by the sublattice model ($La^{3+}$, $Sr^{2+}$, $Mn^{2+}$, $Mn^{3+}$, Y, $Y^{3+}$, Zr, $Zr^{4+}$)$_1$($O^{2-}$, Va)$_2$[21]. The YSZ phase has five independent compositions, e.g., La, Sr, Mn, Y and Zr compositions. Therefore, we can define three auxiliary variables between LSM and YSZ for their common compositions La, Sr and Mn, although the definition range of these compositions may not be the same in the two phases:

$$\phi_1 = X_{La}^{LSM} - X_{La}^{YSZ} \tag{14a}$$

$$\phi_2 = X_{Sr}^{LSM} - X_{Sr}^{YSZ} \tag{14b}$$

$$\phi_3 = X_{Mn}^{LSM} - X_{Mn}^{YSZ} \tag{14c}$$

It should be noted that in addition to these auxiliary composition variables, additional non-conserved order parameters should be defined and evolved for both phases. These additional non-conserved order parameters represent the extents of the internal atomic, electronic and defect processes among sublattice species[19]. The detailed phase-field simulations for these multicomponent multiphase systems will be reported in our forthcoming studies.

## 4. Conclusions

We developed a new model of interfacial compositions by introducing an auxiliary variable to characterize the composition difference between two solution phases. It is applicable to the full range of kinetic conditions from equal interfacial chemical composition to equal diffusion potential and reduced to the WBM and KKS models as two limiting cases. The current model is applicable to kinetic processes involving phases which do not share common composition ranges for their stability, which could pose challenges to the existing WBM and KKS models without doing any simplifications, extrapolations, approximations or fitting to the original chemical potential expressions from thermodynamic database. It is expected that the current model will be widely applied for materials processes involving composition changes as one of the processes.

**Acknowledgement**

The authors acknowledge the partial site contract support from the United States Department of Energy, National Energy Technology Laboratory to support the US Department of Energy's Fossil Energy Solid Oxide Fuel Cell Program, the support from NASA under grant number 80NSSC21M0107, and the Hamer Foundation through the Hamer Professorship at Penn State. The computations were performed on the Roar supercomputer at Pennsylvania State University.

# Supplementary Materials

**S1. About the evolution equation of the auxiliary variable $\phi$, Eq. (8)**

The evolution of $\phi$ should follow the Allen-Cahn equation since it is a non-conserved variable. By directly taking the variational derivative of the total free energy $F$ with respect to $\phi$, we can obtain

$$\frac{\partial \phi}{\partial t} = -L_\phi \frac{\delta F}{\delta \phi} = -L_\phi \left( \frac{1-h(\xi)}{v_m} \frac{\partial \mu^\alpha}{\partial \phi} + \frac{h(\xi)}{v_m} \frac{\partial \mu^\beta}{\partial \phi} \right) \tag{S1-1}$$

This equation can be further simplified. Assuming the three variables, $\xi$, $X_B$ and $\phi$ are independent of each other, then, by taking derivatives with respect to $\phi$ on both sides of Eq. (1) and Eq. (2) in the main text, we have

$$\frac{\partial X_B}{\partial \phi} = \big(1-h(\xi)\big) \frac{\partial X_B^\alpha}{\partial \phi} + h(\xi) \frac{\partial X_B^\beta}{\partial \phi} = 0 \tag{S1-2a}$$

$$\frac{\partial \phi}{\partial \phi} = \frac{\partial X_B^\alpha}{\partial \phi} - \frac{\partial X_B^\beta}{\partial \phi} = 1 \tag{S1-2b}$$

From which we can obtain

$$\frac{\partial X_B^\alpha}{\partial \phi} = h(\xi) \tag{S1-3a}$$

$$\frac{\partial X_B^\beta}{\partial \phi} = -[1-h(\xi)] \tag{S1-3b}$$

Therefore, we can obtain

$$\begin{aligned} \frac{\partial \phi}{\partial t} &= -L_\phi \frac{\delta F}{\delta \phi} = -L_\phi \left( \frac{1-h(\xi)}{V_m} \frac{\partial \mu^\alpha}{\partial \phi} + \frac{h(\xi)}{V_m} \frac{\partial \mu^\beta}{\partial \phi} \right) \\ &= -L_\phi \left( \frac{1-h(\xi)}{V_m} \frac{\partial \mu^\alpha}{\partial X_B^\alpha} \frac{\partial X_B^\alpha}{\partial \phi} + \frac{h(\xi)}{V_m} \frac{\partial \mu^\beta}{\partial X_B^\beta} \frac{\partial X_B^\beta}{\partial \phi} \right) \\ &= -L_\phi \left( \frac{1-h(\xi)}{V_m} \frac{\partial \mu^\alpha}{\partial X_B^\alpha} h(\xi) - \frac{h(\xi)}{V_m} \frac{\partial \mu^\beta}{\partial X_B^\beta} [1-h(\xi)] \right) \\ &= -L_\phi \left( \frac{h(\xi)\big(1-h(\xi)\big)}{V_m} \left( \frac{\partial \mu^\alpha}{\partial X_B^\alpha} - \frac{\partial \mu^\beta}{\partial X_B^\beta} \right) \right) \end{aligned} \tag{S1-4}$$

which is Eq. (8) in the main text.

**S2. A modified Newton-Raphson iterative scheme for solving the phase compositions in the KKS model**

According to the KKS model, the phase compositions are related via

$$X_B = \big(1 - h(\xi)\big)X_B^\alpha + h(\xi)X_B^\beta \quad \text{(S2-1a)}$$

$$\frac{\partial \mu^\alpha}{\partial X_B^\alpha} = \frac{\partial \mu^\beta}{\partial X_B^\beta} \quad \text{(S2-1b)}$$

where $\mu^\alpha$ and $\mu^\beta$ are in general nonlinear functions of $X_B^\alpha$ and $X_B^\beta$, respectively, so that $X_B^\alpha$ and $X_B^\beta$ could not be analytically solved. On the other hand, Eq. (S2-1) can be numerically solved using the Newton-Raphson scheme via

$$\mathbf{x}^{n+1} = \mathbf{x}^n - (\mathbf{J}^n)^{-1}\mathbf{f}^n \quad \text{(S2-2)}$$

where $\mathbf{x} = \left(X_B^\alpha, X_B^\beta\right)^T$ is the unknown vector to be solved,

$$\mathbf{f} = \begin{pmatrix} \big(1 - h(\xi)\big)X_B^\alpha + h(\xi)X_B^\beta - X_B \\ \dfrac{\partial \mu^\alpha}{\partial X_B^\alpha} - \dfrac{\partial \mu^\beta}{\partial X_B^\beta} \end{pmatrix} \quad \text{(S2-3)}$$

is a known vector that should approach **0** upon convergence, and

$$\mathbf{J} = \begin{pmatrix} \big(1 - h(\xi)\big) & h(\xi) \\ \dfrac{\partial^2 \mu^\alpha}{\partial (X_B^\alpha)^2} & -\dfrac{\partial^2 \mu^\beta}{\partial \left(X_B^\beta\right)^2} \end{pmatrix} \quad \text{(S2-4)}$$

is a known coefficient matrix. The superscript *n* indicates the previous iteration step and *n*+1 indicates the new iteration step. To further improve the performance of the iterative scheme, we modify Eq. (S2-2) as

$$\mathbf{x}^{n+1} = \mathbf{x}^n - \lambda(\mathbf{J}^n)^{-1}\mathbf{f}^n \quad \text{(S2-5)}$$

where $\lambda$ is a variable between 0 and 1 so that the function **f** monotonously decreases until reaching the desirable convergence criteria. For each iteration step, we start with an initial guess of $\lambda = 1$; if the $\lambda$ value could not lead to a decrease of **f**, we divide the $\lambda$ by 2.0 every time, until **f** monotonously decreases. For each iteration step, the values of $\mathbf{x}^{n+1}$ are also manipulated to ensure $0 < X_B^\alpha < 1$ and $0 < X_B^\beta < 1$.

### S3. Efficiency test

To further demonstrate the efficiency of the current phase-field model, we consider a 3-D system with 128×128×128 grid points and a grid spacing of 10 nm. The interface thickness is 60 nm. An initial α-Ti nucleus with a radius of 100 nm and composition of $X_V^{\alpha,0} = 0.0323$ was manually introduced at the center of the 3-D system. $\phi$ is evolved using Eq. (8) with a kinetic coefficient $L_\phi = L_\xi$. All the other model parameters are identical to those for the 1-D simulations in Figures 2 and 3. In comparison, the simulation is also performed using the KKS model. Note that the WBM model is not used here for efficiency comparison due to the thick interface and significant extra potential jump at the interface. Our simulation results show that the current phase-field model can guarantee that more than 99.99% of the extra potential is removed at the interface during the evolution, while the simulation is ~3.3 times faster than the KKS model, as shown in Figure S3-1(b) below. This efficiency upgrade is more significant for 3-D systems than 1-D (Figure S3-1(a)) or 2-D systems, since there are more interface points in 3-D systems where more Newton iteration steps are needed for the KKS model.

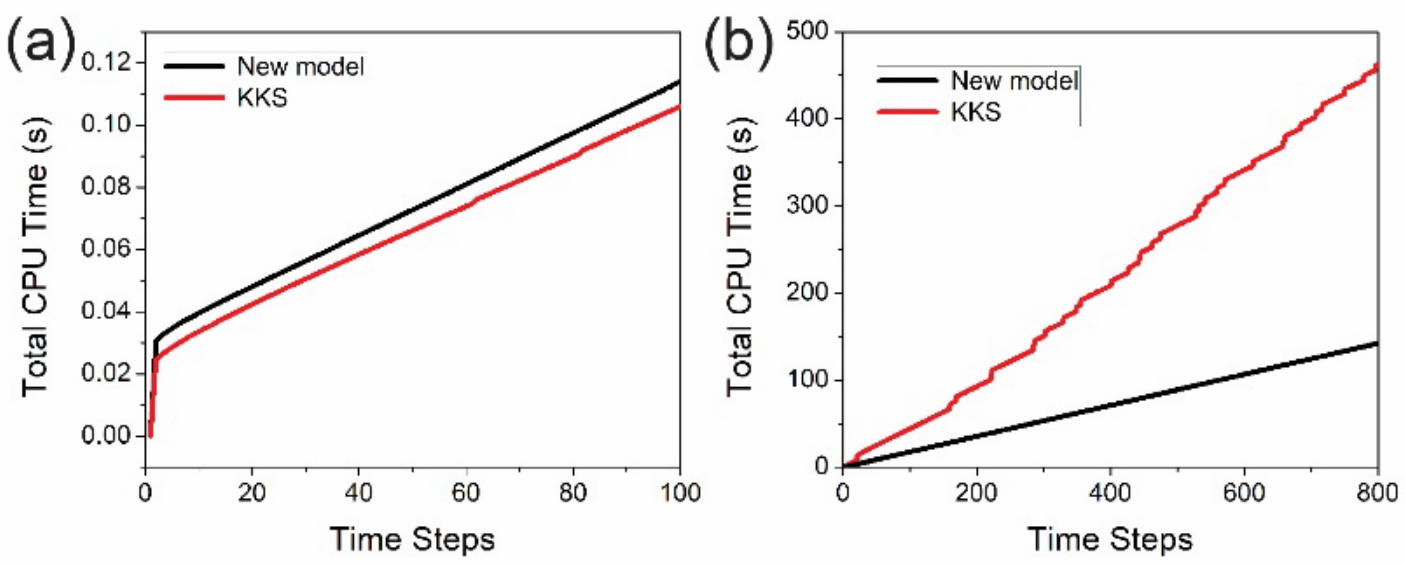

Figure S3-1: Efficiency comparison between the KKS model and the current phase-field models. (a) CPU time required for 1-D simulation with 1024 grid points; (b) CPU time required for 3-D simulation with 128*128*128 grid points.

### S4. Avoiding the logarithm issue: derivation of Eq. (12)

For interdiffusion between nonstoichiometric θ and η1 in Al-Cu alloys, the governing equations for the phase order parameter $\xi$, the composition $X_{Cu}$, and the auxiliary variable $\phi$ should follow Eqs. (7a), (7b) and (8) in the main text, respectively. However, since $\mu^{\theta}$ and $\mu^{\eta 1}$ do not share any common composition range (see Table 2 in the main text), directly solving these equations will be numerically challenging. To be more specific, the expressions of both $\mu^{\theta}$ and $\mu^{\eta 1}$ contain logarithm functions of compositions, which limit the phase composition ranges. Directly evolving Eqs. (7a), (7b) and (8) could not guarantee that the phase compositions are always within their definition domains. The detailed expressions of $\mu^{\theta}$ and $\mu^{\eta 1}$ in terms of phase compositions $X_{Cu}^{\theta}$ and $X_{Cu}^{\eta 1}$, respectively, are:

$$\mu^{\theta} = \frac{1}{3}\Big[\mu_{Al:Al}^{0\theta}\left(1-3X_{Cu}^{\theta}\right) + \mu_{Al:Cu}^{0\theta}\cdot 3X_{Cu}^{\theta} + RT\left(3X_{Cu}^{\theta}\ln 3X_{Cu}^{\theta} + \left(1-3X_{Cu}^{\theta}\right)\ln\left(1-3X_{Cu}^{\theta}\right)\right) + 3X_{Cu}^{\theta}\left(1-3X_{Cu}^{\theta}\right)L^{0\theta}\Big] \quad \text{(S3-1a)}$$

$$\mu^{\eta 1} = \frac{1}{2}\Big[\mu_{Al:Cu}^{0\eta 1}\left(2-2X_{Cu}^{\eta 1}\right) + \mu_{Cu:Cu}^{0\eta 1}\cdot\left(2X_{Cu}^{\eta 1}-1\right) + RT\left(\left(2-2X_{Cu}^{\eta 1}\right)\ln\left(2-2X_{Cu}^{\eta 1}\right) + \left(2X_{Cu}^{\eta 1}-1\right)\ln\left(2X_{Cu}^{\eta 1}-1\right)\right) + \left(2-2X_{Cu}^{\eta 1}\right)\left(2X_{Cu}^{\eta 1}-1\right)L^{0\eta 1}\Big] \quad \text{(S3-1b)}$$

where $\mu_{Al:Al}^{0\theta}$, $\mu_{Al:Cu}^{0\theta}$, $\mu_{Al:Cu}^{0\eta 1}$ and $\mu_{Cu:Cu}^{0\eta 1}$ are standard chemical potentials of endmembers $Al_2Al$ ($\theta$), $Al_2Cu$ ($\theta$), AlCu ($\eta 1$), CuCu ($\eta 1$), respectively; $L^{0\theta}$ and $L^{0\eta 1}$ are interaction parameters, all of which are taken from thermodynamic database[1]. As shown in Eq. (S3-1a), the logarithm term $\ln 3X_{Cu}^{\theta}$ requires $X_{Cu}^{\theta} > 0$ while $\ln\left(1-3X_{Cu}^{\theta}\right)$ requires $X_{Cu}^{\theta} < 1/3$. Therefore, the allowable composition range of $\mu^{\theta}$ is $0 < X_{Cu}^{\theta} < 1/3$. Similarly, according to Eq. (S3-1b), the allowable composition range of $\mu^{\eta 1}$ is $1/2 < X_{Cu}^{\eta 1} < 1$. According to Eq. (1) in the main text, i.e., $X_{Cu} = \big(1-h(\xi)\big)X_{Cu}^{\theta} + h(\xi)X_{Cu}^{\eta1}$, directly evolving $X_{Cu}$ could not guarantee that $X_{Cu}^{\theta}$ and $X_{Cu}^{\eta1}$ are always within their definition range. To overcome this issue, one possible strategy is to find some intermediate variables as a function of the phase compositions, which should be directly evolved with value ranges $(-\infty, +\infty)$ . Then, the phase compositions can be obtained from these variables. We therefore find two such variables $Y_1$ and $Y_2$:

$$Y_1 = \ln\frac{3X_{Cu}^{\theta}}{1-3X_{Cu}^{\theta}} \tag{S3-2a}$$

$$Y_2 = \ln\frac{2X_{Cu}^{\eta1}-1}{2-2X_{Cu}^{\eta1}} \tag{S3-2b}$$

According to Eqs. (1-2) in the main text, these variables can be further expressed in terms of the phase order parameter $\xi$, the local composition $X_{Cu}$ and the auxiliary variable $\phi$ as

$$Y_1 = \ln\frac{3[X_{Cu}+h(\xi)\phi]}{1-3[X_{Cu}+h(\xi)\phi]} \tag{S3-3a}$$

$$Y_2 = \ln\frac{2\big[X_{Cu}-\big(1-h(\xi)\big)\phi\big]-1}{2-2\big[X_{Cu}-\big(1-h(\xi)\big)\phi\big]} \tag{S3-3b}$$

From which we can obtain the expressions of $X_{Cu}$ and $\phi$ as a function of $\xi$, $Y_1$ and $Y_2$:

$$X_{Cu} = \big(1-h(\xi)\big)\frac{e^{Y_1}}{3+3e^{Y_1}} + h(\xi)\frac{2e^{Y_2}+1}{2+2e^{Y_2}} \tag{S3-4a}$$

$$\phi = \frac{e^{Y_1}}{3+3e^{Y_1}} - \frac{2e^{Y_2}+1}{2+2e^{Y_2}} \tag{S3-4b}$$

Plugging Eq. (S3-4a) into the left-hand side of Eq. (7b) in the main text (i.e., $\frac{\partial X_{Cu}}{\partial t}$) and plugging Eq. (S3-4b) into the left-hand-side of Eq. (8) in the main text (i.e., $\frac{\partial \phi}{\partial t}$), we can obtain

$$\frac{\partial X_{Cu}}{\partial t} = \big(1-h(\xi)\big)\frac{3e^{Y_1}}{(3+3e^{Y_1})^2}\frac{\partial Y_1}{\partial t} + h(\xi)\frac{2e^{Y_2}}{(2+2e^{Y_2})^2}\frac{\partial Y_2}{\partial t} + \frac{\partial h}{\partial \xi}\frac{\partial \xi}{\partial t}\left(\frac{2e^{Y_2}+1}{2+2e^{Y_2}} - \frac{e^{Y_1}}{3+3e^{Y_1}}\right) \tag{S3-5a}$$

$$\frac{\partial \phi}{\partial t} = \frac{3e^{Y_1}}{(3+3e^{Y_1})^2}\frac{\partial Y_1}{\partial t} - \frac{2e^{Y_2}}{(2+2e^{Y_2})^2}\frac{\partial Y_2}{\partial t} \tag{S3-5b}$$

which can be further arranged as (note that $\frac{2e^{Y_2}+1}{2+2e^{Y_2}} - \frac{e^{Y_1}}{3+3e^{Y_1}} = -\phi$)

$$\frac{3e^{Y_1}}{(3+3e^{Y_1})^2}\frac{\partial Y_1}{\partial t} = \frac{\partial X_{Cu}}{\partial t} + h(\xi)\frac{\partial \phi}{\partial t} - \phi\frac{\partial h}{\partial \xi}\frac{\partial \xi}{\partial t} \tag{S3-6a}$$

$$\frac{2e^{Y_2}}{(2+2e^{Y_2})^2}\frac{\partial Y_2}{\partial t} = \frac{\partial X_{Cu}}{\partial t} - \big(1-h(\xi)\big)\frac{\partial \phi}{\partial t} - \phi\frac{\partial h}{\partial \xi}\frac{\partial \xi}{\partial t} \tag{S3-6b}$$

Replacing $\frac{\partial X_{Cu}}{\partial t}$ in these equations with the right-hand side of Eq. (7b), replacing $\frac{\partial \phi}{\partial t}$ in these equations with the right-hand side of Eq. (8) and replacing $\frac{\partial \xi}{\partial t}$ in these equations with the right-hand side of Eq. (7a), we can obtain the governing evolution equations for the intermediate variables $Y_1$ and $Y_2$:

$$\frac{3e^{Y_1}}{(3+3e^{Y_1})^2}\frac{\partial Y_1}{\partial t} = \nabla\cdot M_{Cu}\nabla\frac{\delta F}{\delta X_{Cu}} - h(\xi)\frac{L_\phi}{v_m}\left(\frac{\partial\mu^\theta}{\partial X_{Cu}^\theta} - \frac{\partial\mu^{\eta 1}}{\partial X_{Cu}^{\eta 1}}\right) + \phi\frac{\partial h}{\partial\xi}L_\xi\frac{\delta F}{\delta\xi} \quad \text{(S3-7a)}$$

$$\frac{2e^{Y_2}}{(2+2e^{Y_2})^2}\frac{\partial Y_2}{\partial t} = \nabla\cdot M_{Cu}\nabla\frac{\delta F}{\delta X_{Cu}} + \big(1-h(\xi)\big)\frac{L_\phi}{v_m}\left(\frac{\partial\mu^\theta}{\partial X_{Cu}^\theta} - \frac{\partial\mu^{\eta 1}}{\partial X_{Cu}^{\eta 1}}\right) + \phi\frac{\partial h}{\partial\xi}L_\xi\frac{\delta F}{\delta\xi} \quad \text{(S3-7b)}$$

which are Eq. (12a) and Eq. (12b) in the main text. Therefore, now we are evolving Eq. (7a), Eq. (S3-7a) and Eq. (S3-7b) to obtain ξ, $Y_1$ and $Y_2$ values. The $Y_1$ and $Y_2$ values can then be taken into Eqs. (S3-4a) and (S3-4b) to obtain $X_{Cu}$ and $\phi$ values, and taken into

$$X_{Cu}^\theta = \frac{e^{Y_1}}{3+3e^{Y_1}} \quad \text{(S3-8a)}$$

$$X_{Cu}^{\eta 1} = \frac{2e^{Y_2}+1}{2+2e^{Y_2}} \quad \text{(S3-8b)}$$

to obtain the phase compositions. In this way, the phase compositions $X_{Cu}^\theta$ and $X_{Cu}^{\eta 1}$ would never get out of their respective ranges of $0 < X_{Cu}^\theta < \frac{1}{3}$ and $\frac{1}{2} < X_{Cu}^{\eta 1} < 1$, and the possible numerical issue due to the logarithm expressions is resolved.